\documentclass[prb,twocolumn,amsmath,amssymb]{revtex4}
\usepackage{graphicx} 

\newcommand{\ddgb}{\Delta\Delta G_\text{bind}}
\newcommand{\Ur}{U_\text{r}}
\newcommand{\pmf}{\Delta G}
\newcommand{\dS}{\Delta S}
\newcommand{\dH}{\Delta H}

\begin{document}

\title{Computational study of small molecule binding
for both tethered and free conditions}
\author{F.\ Marty Ytreberg\footnote{ytreberg@uidaho.edu}}
\affiliation{Department of Physics, University of Idaho, Moscow, ID 83844-0903}
\date{\today}

\begin{abstract}
Using a calix[4]arene-benzene complex as a test system
we compare the potential of mean force for when the calix[4]arene
is tethered versus free.
When the complex is in vacuum our results show that the difference
between tethered and free is primarily due to the entropic
contribution to the potential of mean force resulting in a binding free
energy difference of 6.5 kJ/mol.
By contrast, when the complex is in water our results suggest that
the difference between tethered and free is due to the enthalpic
contribution resulting in a binding free energy difference of 1.6 kJ/mol.
This study elucidates the roles of entropy and enthalpy for this small
molecule system and emphasizes the point that tethering the receptor
has the potential to dramatically impact the binding properties.
These findings should be taken into consideration when using
calixarene molecules in nanosensor design.
\end{abstract}

\maketitle

\section{Introduction}

Calixarenes are macrocycles that are of interest due to the fact that they can
be easily synthesized and can be functionalized to selectively
bind neutral or ionic analytes; see review refs
\onlinecite{denamor-zapata,schatz,ludwig,princy-shobana,sameni-harrowfield}.
One use for calixarenes that is of specific interest to the
current study is in nanosensor design
(e.g., refs
\onlinecite{filenko-rangelow,koshets-kalchenko,chen-xu,dickert-schuster}).
Calixarenes are typically used in nanosensors by
decorating the nanomaterial with gold
and then tethering the calixarenes to the gold surface.
A reasonable question that is also the motivation for our study is:
How does tethering calixarene to a surface affect the
binding properties of the calixarene to analytes?
This is an important question considering that typically a researcher
may only have knowledge of binding properties for free
(i.e., not tethered) conditions.

For the current study we compare the effects of tethering a calix[4]arene
on the binding properties for both in vacuum and in water.
Note that the ``[4]'' means that there are four aromatic
rings in the structure.
To our knowledge there are no previous studies that have determined
the affects of such tethering on the binding properties of calixarenes.
Thus, we computed the potential of mean force (PMF)
for calix[4]arene-benzene binding for four cases:
(i) in vacuum with the calix[4]arene tethered;
(ii) in vacuum with the calix[4]arene free (i.e., not tethered);
(iii) in water with the calix[4]arene tethered; and
(iv) in water with the calix[4]arene free.
Our results below show that when the complex is in vacuum
the difference between tethered and free is due primarily to the entropic
contribution to the potential of mean force resulting in a binding free
energy difference of 6.5 kJ/mol.
By contrast, when the complex is in water our results suggest that
the difference between tethered and free is due entirely to the enthalpic
contribution resulting in a binding free energy difference of 1.6 kJ/mol.

\section{Computational Methods}

The initial structure for the calix[4]arene-benzene complex
was obtained from experimental X-ray crystallography
(personal communication from Pam Shapiro's lab at University
of Idaho).
The necessary simulation topologies for both the calix[4]arene and benzene
were then generated by the PRODRG server \cite{prodrg}.
We then modified the partial charges to be consistent with
the GROMOS-96 43A1 forcefield \cite{gromos}, e.g., all CH3 
groups were set to zero partial charge.
The GROMACS simulation package version 3.3.3 \cite{gromacs}
was used for all molecular dynamics simulations described below
with the default GROMOS-96 43A1 forcefield \cite{gromos}.

For the vacuum simulations the calix[4]arene-benzene complex was
first minimized using steepest decent for 1000 steps.
For subsequent production simulations all Van der Waals and
electrostatic interactions were computed, i.e., no cutoffs were used.
A timestep of 1.0 fs was utilized with no constraints. 
The temperature was maintained at a constant value using
Langevin dynamics \cite{langevin} with a friction coefficient of 1.0 amu/ps.

For the simulations in water the calix[4]arene-benzene complex
was solvated in a cubic box of SPC water \cite{spc} of
approximate initial size 4.5 nm a side.
The system was then minimized using steepest decent for 1000 steps.
To allow for some equilibration of the water the system was then simulated
for 100 ps with the positions of all heavy atoms in the complex
harmonically restrained with a force constant of 1000 kJ/mol/$\text{nm}^2$.
For this equilibration simulation the pressure was maintained at
1.0 atm using the Berendsen algorithm \cite{berendsen}.
Subsequent production simulations were carried out
with the volume fixed at the final value from the equilibration.
For all water simulations the LINCS algorithm \cite{lincs}
was used to constrain hydrogens to their ideal lengths allowing the
use of a 2.0 fs timestep.
The temperature was maintained at a constant value using
Langevin dynamics \cite{langevin} with a friction coefficient of 1.0 amu/ps.
Particle mesh Ewald \cite{pme}
was used for electrostatics with a real-space cutoff of 1.0 nm
and a Fourier spacing of 0.1 nm.  Van der Waals interactions used a
cutoff with a smoothing function such that the interactions smoothly
decayed to zero between 0.75 nm and 0.9 nm.
Dispersion corrections for the energy and pressure were utilized
\cite{allen-tildesley}.

\begin{figure*}
    \begin{center}
	\includegraphics[height=2.4in]{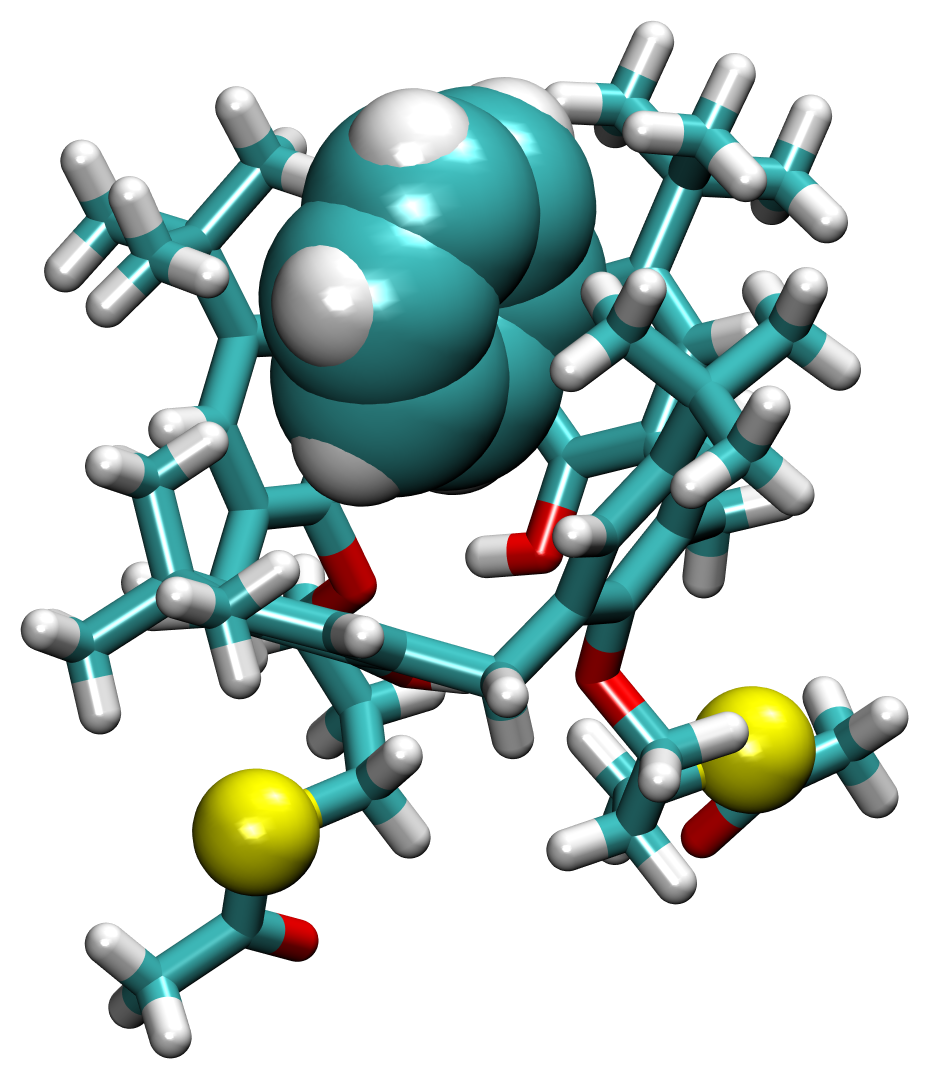}
    \end{center}
    \caption{ \label{fig-calix}
    The calix[4]arene-benzene model system used for the current study.
    The calix[4]arene molecule has a basket shaped binding pocket
    and can be functionalized to bind both neutral and ionic analytes.
    The two sulfur atoms are shown in a larger size and allow the
    calix[4]arene to be tethered to a gold surface.
    The difference between the tethered and free simulations
    in the current study is that these sulfur atoms were
    harmonically restrained to the position shown in the figure during the
    tethered simulations
    but were not restrained for the free simulations.
    This image was generated using VMD \cite{vmd}.
    }
\end{figure*}

To perform the tethered simulations for both vacuum and in water
we harmonically restrained the two sulfur atoms
shown in \ref{fig-calix}
using a force constant of 10,000 kJ/mol/$\text{nm}^2$.
The purpose is to mimic the effect of the calix[4]arene binding
to a gold surface.
This harmonic restraint on the sulfur atoms
was not present for the free simulations.

\subsection{Generating PMF estimates}

We computed all PMFs using umbrella sampling and
weighted histogram analysis (WHAM) \cite{swendsen-wham}.
Our technique for estimating the PMF using WHAM is described
in ref \onlinecite{ytreberg-fkbp}.
Briefly, the GROMACS 3.3.3 software package \cite{gromacs}
was modified to provide a harmonic biasing potential
$\Ur(r)=0.5k_r(r-r_0)^2$ which depends only on the center of
mass separation $r$ between the calix[4]arene and the benzene.
For all PMF estimates we used a total of 33 windows
$r_0=0.40,0.45,0.50,\ldots,1.95,2.00$.
For the vacuum system each window was simulated for 32.0 ns;
16.0 ns were discarded for equilibration and 16.0 ns were used for
the WHAM analysis.
For the water system each window was simulated for 4.0 ns;
2.0 ns were discarded for equilibration and 2.0 ns were used for
the analysis.
For all PMF estimates below the biasing potential $\Ur$
used a force constant $k_r=3000$ kJ/mol/$\text{nm}^2$
and the estimates include the $2\ln(r)$ Jacobian correction
\cite{ytreberg-fkbp,vangunsteren-pmf}.

Note that for the simulations of the complex in water the system size
prevents the long simulation times necessary to obtain converged PMFs
without additional restraints.
Thus, for the water simulations (but not for the vacuum simulations)
we utilized an axial restraint that keeps the 
benzene on the binding axis relative to the calix[4]arene
as described in ref \onlinecite{ytreberg-fkbp}.
Use of this restraint means that it is not valid to directly compare
the vacuum and water PMFs.
However, it is still valid to compare the tethered
and free conditions for water
which is the purpose of this study.

\subsection{Estimating entropic and enthalpic contributions}

To estimate the entropic contribution to the PMF $T\dS(r)$
we used the fact that the entropy is related to the derivative of
the PMF $\pmf(r)$ with respect to system temperature $T$
(see also refs \onlinecite{ghosh-garcia,choudhury-pettitt}),
\begin{equation}
    T\dS(r)=-T\bigg( \frac{\partial \pmf(r,T)}{\partial T} \bigg)\,.
    \label{eq-TS}
\end{equation}
This derivative was numerically estimated by computing the PMF at
three temperatures 270 K, 300 K and 330 K
and then using a three-point finite difference approximation.
The enthalpic contribution $\dH(r)$ was then estimated via
\begin{equation}
    \dH(r)=\pmf(r)+T\dS(r).
    \label{eq-H}
\end{equation}

\subsection{Uncertainty estimation}

The uncertainty for $\pmf$, $T\dS$ and $\dH$ were estimated 
by computing the standard error over independent trials.
For both the tethered and free conditions in vacuum 10 independent
estimates of the PMF were generated at each of the three temperatures
(i.e., 30 PMF estimates tethered and 30 free).
For both the tethered and free conditions in water five independent
estimates of the PMF were generated at each of the three temperatures
(i.e., 15 PMF estimates tethered and 15 free).

\section{Results and Discussion}

\begin{figure*}
    \begin{center}
	\includegraphics[height=2.4in]{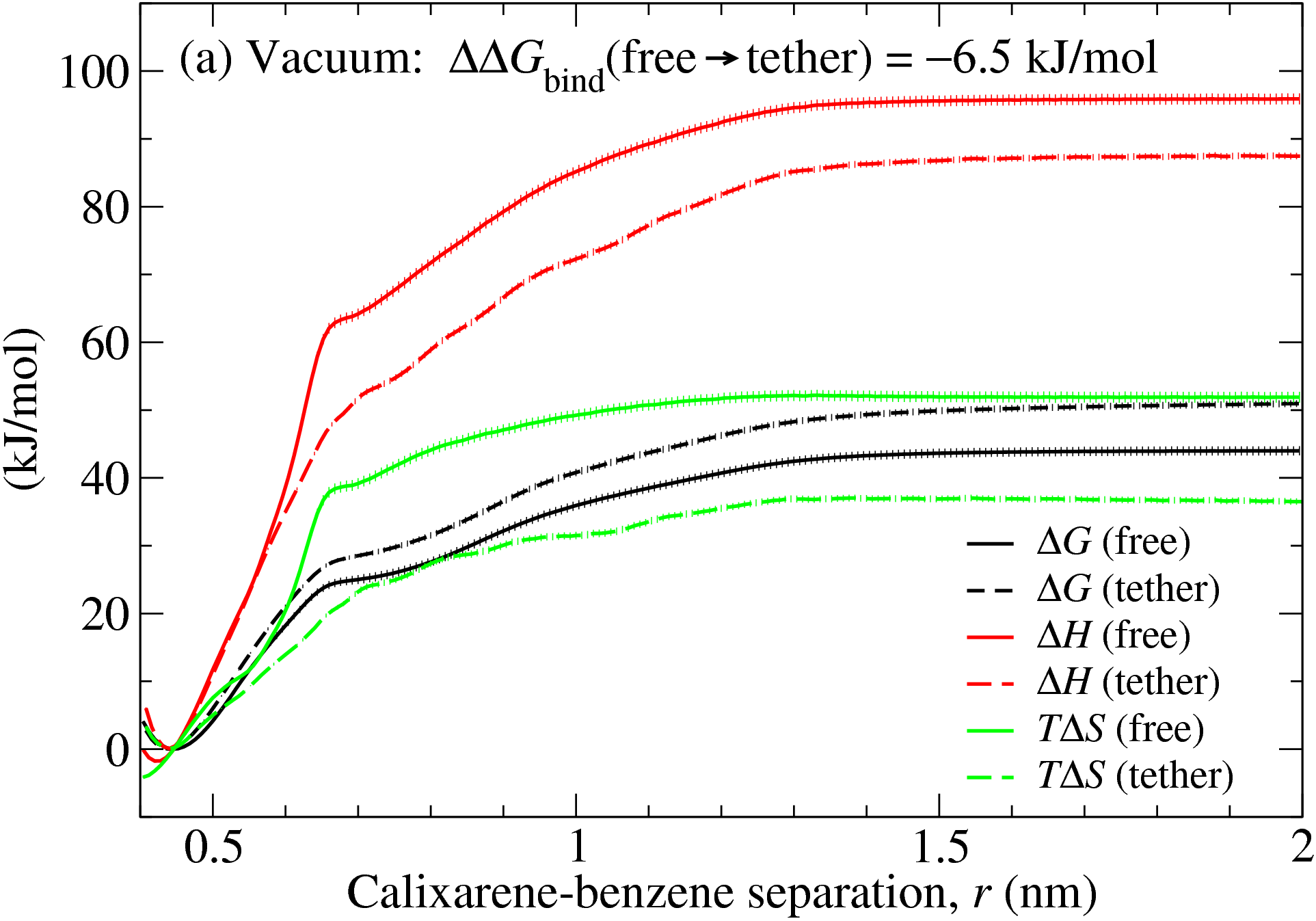}
	\hfill
	\includegraphics[height=2.4in]{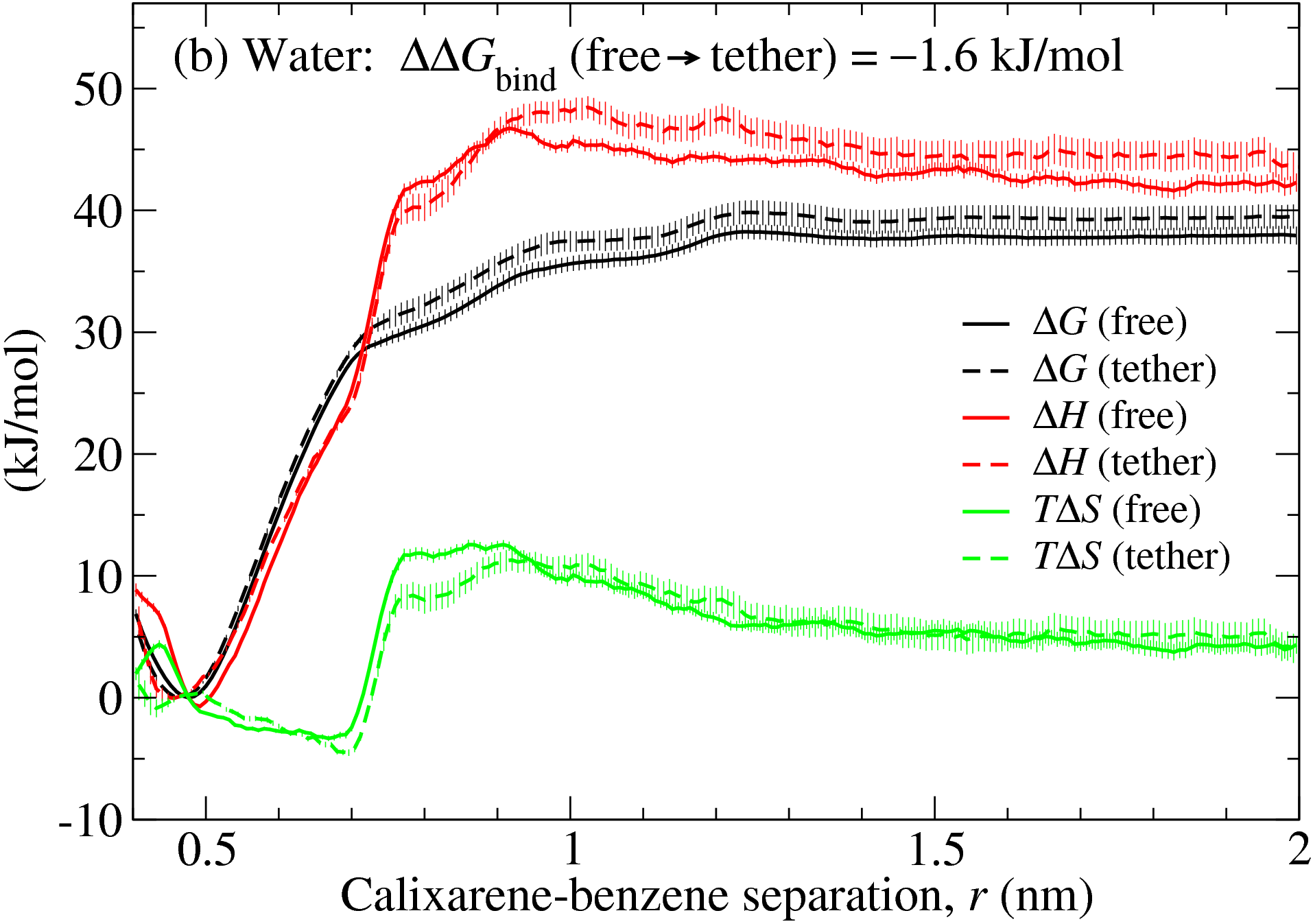}
    \end{center}
    \caption{ \label{fig-pmf}
    The calix[4]arene-benzene potential of mean force (black)
    showing the enthalpic (red) and entropic (green) contributions.
    Both tethered (dashed line) and free (solid line) conditions are shown.
    The error bars are the standard error obtained from performing
    multiple independent simulations.
    (a) Simulation results in vacuum.
    Due primarily to the entropic contribution there is a free energy
    difference of 6.5 kJ/mol between tethered and free conditions.
    (b) Simulation results in water.
    Due primarily to the enthalpic contribution there is a free energy
    difference of 1.6 kJ/mol between tethered and free conditions.
    }
\end{figure*}

Results for both vacuum and water simulations are shown in
\ref{fig-pmf}.
The binding free energy differences between
tethered and free conditions $\ddgb$ for vacuum and water
were obtained by numerically
integrating the $\pmf(r)$ curves from $r=0.4$ to $r=1.9$ nm.
The entropic and enthalpic contributions to the
PMF were computed using eqs \eqref{eq-TS} and \eqref{eq-H}.

Results for the calix[4]arene-benzene complex
in vacuum are shown in \ref{fig-pmf}a
and reveal two major differences between the free and tethered conditions.
First, when the PMF plateaus ($r > 1.5$ nm)
both the entropic and enthalpic contributions
for the free conditions is larger than for tethered.
Note that the free entropic contribution is larger than tethered
by approximately 15 kJ/mol, and the enthalpic contribution is larger
by about 8 kJ/mol.
Thus, it is primarily the entropic difference between free and tethered
that leads to the more favorable binding for tethered conditions.
The second difference between free and tethered conditions
occurs at $r \approx 0.6$ nm where the entropic and enthalpic
contributions to the PMF increase dramatically
for free as compared to tethered conditions.
This difference can be understood by noting that the benzene is 
at the (wider) outer edge of the binding pocket at $r \approx 0.6$ nm;
see \ref{fig-calix}.
Benzene is just outside the binding pocket at $r \approx 0.7$ nm
where entropic and enthalpic contributions
for the free condition have a narrow flat region.
We believe that this difference between the free and tethered 
can be attributed to the fact that the tethering of the calix[4]arene
provides a more rigid binding pocket than under free conditions
leading to a sharp increase in entropy under free conditions as
the benzene reaches the outer edge of the pocket.

Our vacuum results suggest that the free energy of binding
under tethered conditions is more favorable than free by $\ddgb=-6.5$ kJ/mol,
due primarily to the entropic contribution to the PMF.
Thus, if one wishes to design a gas phase nanosensor using calix[4]arenes
we strongly suggest testing the binding properties under tethered conditions.

Results for the calix[4]arene-benzene complex
in water are shown in \ref{fig-pmf}b.
The only appreciable difference between tethered and free
conditions is the enthalpic contribution when the PMF plateaus ($r>1.5$ nm).
Interestingly, there is no difference (within error)
between the free and tethered entropic contributions to the PMF
for $r>1.5$ nm.
Apparently the entropy of the water molecules completely counters
the entropy-dominant effects seen in the vacuum system.
We believe this is attributed to the fact that the effective volume
available to the water molecules is reduced when the benzene is completely
dissociated from the calix[4]arene; this also leads to the entropy decrease
observed in the PMFs for both tethered and free between
$r\approx 0.9$ and $r\approx 1.5$ nm.

Our results in water suggest that the free energy of binding
under tethered conditions is more favorable than free by $\ddgb=-1.6$ kJ/mol.
In contrast to vacuum this free energy difference
is due entirely to the enthalpic contribution to the PMF.

Note that we do not expect that our results are completely general
and thus different receptor-compound
complexes will likely lead to differences from our results above.
However, this does not change our general conclusion that one must
test binding properties under the desired conditions.

\section{Conclusion}

We have studied the effects of tethering on small molecule binding properties
using a calix[4]arene-benzene complex as a test system.
Simulations of the complex in vacuum and in water were performed
and the potential of mean force (PMF) curves were computed and compared
for tethered and free conditions.

Our results for the calix[4]arene-benzene complex in vacuum show
that the primary difference between free and tethered conditions
is the entropic contribution to the PMF.
Thus, in vacuum the free energy of binding under tethered conditions
is more favorable than free by $\ddgb=-6.5$ kJ/mol.
By contrast, when the calix[4]arene complex is in water
the only substantial difference between free and tethered
conditions is the enthalpic contribution to the PMF.
Thus, in water the free energy of binding under tethered conditions
is more favorable than free by $\ddgb=-1.6$ kJ/mol.

This study elucidates the roles of entropy and enthalpy under tethered
and free conditions for both vacuum and in water.
Our results show substantial differences
in binding properties between tethered and free conditions,
especially in vacuum.
Thus, if one wishes to design a gas phase
or aqueous nanosensor using calix[4]arenes
we suggest that the binding properties of the calix[4]arene
should be tested under tethered conditions.

\section*{Acknowledgements}
The author thanks Pam Shapiro and Steven Hung for providing
the experimental structure for the calix[4]arene-benzene complex,
and Conrad Shyu for helpful discussion.
The project described was supported by Award Numbers P20RR016448
and R21GM083827 from the National Institutes of Health.
The content is soley the responsibility of the authors and does
not necessarily represent the official views of the National
Institutes of Health.
The research was also supported by Idaho NSF-EPSCoR,
and by IBEST and BANTech at University of Idaho.

\bibliography{/home/marty/texmf/bibtex/fmy}

\end{document}